\documentclass[10pt,reprint]{cqos}

\usepackage{amsmath}
\usepackage{mathptmx}
\usepackage[pdftex]{graphicx}
\usepackage{subfigure}

\begin{document}

\setlength{\pdfpageheight}{\paperheight}
\setlength{\pdfpagewidth}{\paperwidth}

\title{QoS-Based Pricing and Scheduling of Batch Jobs in OpenStack Clouds}

\authorinfo{Thomas Sandholm, Julie Ward, Filippo Balestrieri and Bernardo A. Huberman}
           {HP Labs, Palo Alto, CA, USA}
           {\{thomas.e.sandholm,jward,filippo.balestrieri,bernardo.huberman\}@hp.com}
\maketitle

\begin{abstract}
The current Cloud infrastructure services (IaaS) market employs a resource-based selling model: customers rent nodes from the provider and pay per-node per-unit-time.
This selling model places the burden upon customers to predict their job resource requirements and durations. Inaccurate prediction by customers can result in over-provisioning of resources, or under-provisioning and poor job performance.
Thanks to improved resource virtualization and multi-tenant performance isolation, as well as common frameworks for batch jobs, such as MapReduce, Cloud providers can  predict job completion times more accurately.
We offer a new definition of QoS-levels in terms of job completion times and we present a new QoS-based selling mechanism for batch jobs in a multi-tenant OpenStack cluster.
Our experiments show that the QoS-based solution yields up to 40\% improvement over the revenue of more standard selling mechanisms based on a fixed per-node price across various demand and supply conditions in a 240-VCPU OpenStack cluster.
\end{abstract}



\section{Introduction}
Cloud computing is increasing in popularity as networking and
virtualization technology advancements make the move away from
proprietary platforms more beneficial. During the early days of
the Cloud, the typical application was a Web server, run in the Cloud by a start-up to avoid up-front investment in hardware when
future demand was uncertain. Today many types of applications
run in the Cloud, including consumer desktop applications,
such as word processing, enterprise customer relationship management,
and compute- and IO-intensive batch applications. This last kind of application is the focus of our work.

Most Cloud resources are sold as infrastructure services (IaaS) on a per-node-hour
basis and providers allocate nodes on a first-come-first-serve basis to incoming jobs. This model has shortcomings when applied to batch applications. When launching batch jobs, the user is often concerned about his job completion times. However, current business models require the customer to estimate the exact amount of resources needed to complete the job by a given time, which may be difficult. 
For example, the user cannot predict load-dependent interference from other jobs hosted in the same data center.

In this paper we present a system based on a new, results-based selling model for Cloud infrastructure services. In the new model, QoS levels are redefined in terms of job completion times. The Cloud provider offers the user different completion-times at different prices. We show that our solution presents several advantages compared to more standard per-node-hour based selling mechanisms.

For instance, the strategy of letting the user add new resources over time is not efficient from the provider's prospective, since the data may have to
be rebalanced across nodes to ensure that computations take place close to the data. Hence, the fact that the Cloud provider controls the node allocation over time is a desirable feature of our solution.
Secondly, in current resource-based mechanisms, the distribution of nodes to jobs is often based on a first-come-first-serve rule. In other terms, there is no way for a user to signal the importance of his\footnote{We refer to a buyer of resources as {\it he} and a seller as {\it she}} jobs relative to other users' jobs that were submitted earlier.
This may create situations in which the provider forgoes the opportunity to increase her profits. By offering differently priced completion times, instead, the provider is able to price discriminate among customers with different job urgencies. 

To better respond to the user's needs, some Platform-as-a-Service (PaaS) offerings have auto-scale capabilities, i.e. they automatically increase (or decrease) the number of nodes assigned to a job whenever the user increases (decreases) his activity.~\footnote{http://aws.amazon.com/elasticbeanstalk/}
In some cases~\footnote{https://cloud.google.com/appengine/}, the service allows the user to fix a budget. This puts a limit on the extent at which the system can scale up. 
However, these adjustments to the node allocations are based exclusively on the individual user's behavior and not on the overall market dynamics, which is what the provider would prefer and our solution offers.
Spot-market models~\footnote{http://aws.amazon.com/ec2/purchasing-options/spot-instances/} are one exception in the space of resource-based selling mechanisms. Indeed, in a Spot-market, the per-node price varies according to the overall dynamics of supply and demand across all markets of cloud instances. Still, the Spot market may not be the most convenient solution for users: first, the user's expense remains uncertain until job-completion; second, the user may lose his instances at any time and no guarantees are given in terms of completion time.

Our results-based mechanism requires coordination between different components: job completion-time prediction, pricing, and scheduling. In this paper we focus on the integration of these different elements with the OpenStack data processing system. Furthermore, we run experiments that show that we can increase the revenue by 40\% compared to a fixed, per-node-period pricing method.

The rest of the paper is organized as follows. In Section~\ref{sec:related} we review related work.
Section~\ref{sec:pricemodel} outline our QoS model.
In Section~\ref{sec:implementation} we describe the architecture of our systems
integration with OpenStack.
We present the experiment and our key findings in Section~\ref{sec:experiments}
and conclude in
Section~\ref{sec:conclusions}.

\section{Related Work}\label{sec:related}

Related work can be categorized into four main areas: 
(1) resource-based pricing and resource-allocation models for the Cloud;
(2) systems for predicting workload resource requirements and completion times;
(3) systems for results-based resource allocation in the cloud in the absence of pricing;
and (4) results-based pricing and resource-allocation models for cloud services.

Most work in pricing of cloud services is based on resource-based selling models, in which customers pay per node-hour.
Resource-based mechanisms closely resemble the most common selling practices in the current public Cloud service market
and have been the focus of the growing revenue management literature applied to IT resources (\cite{dube2005yield, sulistio2007using, anandasivam2009managing,meinl2010}).
Within this class of mechanisms, the Cloud service provider's pricing and resource allocation problem represents a new area of application for yield management. The solutions derived to maximize revenue from the sale of airline tickets or hotel rooms, such as fare classes with booking limits, can be extended to the new context of computing nodes.

Quasar~\cite{delimitrou2014} is a system that allows workload prediction in the Cloud based on
a novel benchmarking and forecasting model. Their approach involves running micro-benchmarks
of jobs and then utilizing techniques from collaborative filtering algorithms to infer
job completion times at different resource levels. As with our work, their focus is on
batch applications.
The main difference between their work and ours is that we also
consider the pricing of QoS-based contracts. In~\cite{zhang2014} the authors propose a system that offers deadline guarantees for
MapReduce jobs and observe that picking the correct type of resources, a few high-cpu, high-memory
nodes or a high number of low-cpu, low-memory nodes, can
result in vastly different completion times even when the cost is the same. Hence, they propose a predictor that also
determines which resource type is optimal for a job.
Again, no pricing model is proposed in that
work; it relies on the Amazon EC2 on-demand pricing model.

Some papers consider results-based resource allocation without the pricing component. For example, ~\cite{verma2012} and ~\cite{wolf2010} offer deadline-based guarantees in a MapReduce setting by modifying
the way the Hadoop scheduler picks tasks from queues.  Although our focus in this paper is on Hadoop MapReduce
workloads we want to support arbitrary batch applications and thus we do not modify the
Hadoop scheduler, but instead make all the scheduling decisions on a VM or virtual cluster
allocation level. The advantage of this is that we can isolate the performance and data more
strictly, which is more appropriate for public cloud offerings, or cases where sensitive data
are processed. ~\cite{sandholm2009} explored this VM scheduling technique in terms of per-VM
resource configuration. Here we partition resources in terms of whole
virtual clusters instead, as most public clouds today do not allow increasing or decreasing
the resource consumption on an individual VM once allocated. In terms of batch job scheduling in
general a lot of focus has been on fair allocations in private or institutional Clouds, e.g.
~\cite{zaharia2010,isard2009}.

Recently, a number of efforts have also been made to provide results-based pricing schemes for the Cloud (\cite{shi2014,wang2013,qiu2013,jain2014}). While these papers acknowledge 
users' differing sensitivity to job completion time, they rely on the user's truthful revelation of the job resource requirements. Moreover, their objective is efficiency. Given that, they focus on the retrieval of each user's willingness-to-pay for different completion times. The Cloud provider learns such information either assuming that users reveal it truthfully or running properly designed auction-like mechanisms. 
In our work, the Cloud provider determines autonomously each job's completion time. Moreover, our objective is profit maximization. Ergo, we are not interested in learning each user's willingness to pay\footnote{only the general distribution for different broader customer and job types} at any (informational) cost.

\section{QoS-Based Selling Mechanism}\label{sec:pricemodel}
We consider a Cloud provider serving multiple users who submit their jobs over time.~\footnote{The setting implies an on-line scheduling problem} Each user is privately informed about his willingness-to-pay for the job. We assume that such willingness-to-pay is a decreasing function of the job completion-time. Considering a given time $t$, the interaction between users and cloud provider can be described as a multi-step process.
1) At time $t$ a number of jobs $j$ are submitted by different users. 
2) Once a job is submitted, the cloud provider determines a relation that maps the number of nodes allocated to the job to completion times. 
3) The cloud provider offers a menu in which each item is a combination of a price and a job-completion-time.  
4) Given the menu offered by the cloud provider and his willingness to pay, the user decides whether to select an item from the menu and sign a contract with the cloud provider.

\section{OpenStack Integration}\label{sec:implementation}
We have implemented the proposed QoS model for the OpenStack
Juno distribution. The key point of integration is the OpenStack Sahara
data processing API. Sahara allows virtual clusters to be created and
provisioned with the required batch processing software such as
Hadoop, Spark, Oozie and Hive. It is responsible for the optimal placement,
mapping onto physical nodes,
of the individual nodes based on the type of node, e.g., whether it is a
Hadoop master node or a worker node. It also ensures that true data parallelism
and reliability are achieved with anti-affinity rules.
For instance, two Hadoop data nodes should not be allocated on the same physical node
since a physical node failure would take out more than one replica. Hence, in our
implementation we are not concerned with physical node placement and simply
try to determine how large of a dedicated virtual cluster to allocate to a job, and then rely on
Sahara to make the optimal placement decisions. We do, however, configure node templates
and anti-affinity rules to give Sahara hints as to where to place nodes.
We also use templates to distinguish which nodes may be subject to flexing or scaling.
For example, in a Hadoop cluster, it is desirable to keep some nodes available at all times to
make sure there is not too much re-balancing overhead when trying to re-replicate
data to compensate for nodes that went down. Similarly scaling up or down
master or coordinator nodes would not make much sense.

The second Sahara feature we leverage is the Job execution abstraction.
Again, by using templates many different types of jobs may be executed and monitored in a consistent
way. As long as a job is fully specified and configured to be executed within
the Sahara job framework, we have all the information we need to offer a QoS and
pricing contract.
Sahara may take care of Swift input and output staging, Oozie workload
orchestration and Pig script execution for example.

We pre-configure node and cluster templates for each QoS level we offer.
Separate templates may also be configured for prediction benchmarking.
Compared to traditional use of Sahara, once the job is configured and ready to
launch there is no need to configure and create a cluster to run on in our system.
The user simply requests a quote for a job and we then generate a completion time
and price
menu from which they can choose when submitting their job.

\begin{figure}[htp]
	\begin{center}
\includegraphics[scale=0.42]{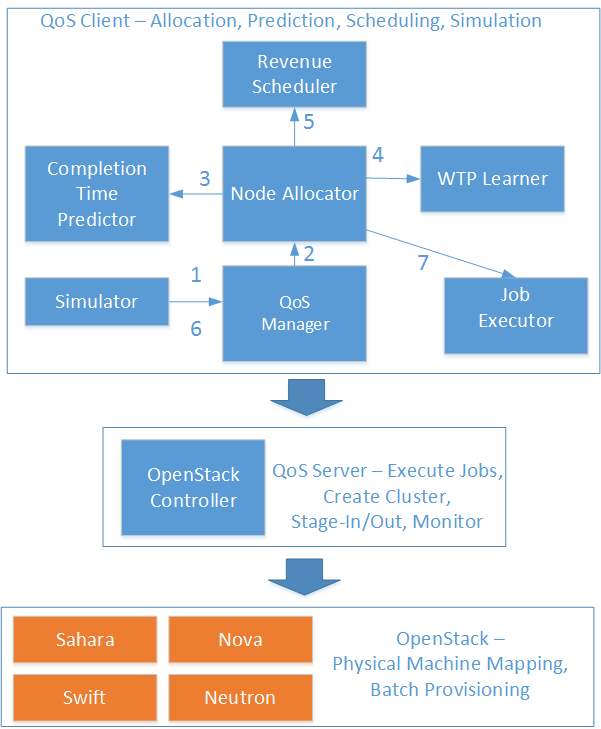}
\caption{Integration Design.
     Price-aware scheduling design in OpenStack.} \label{fig:architecture}
	\end{center}
\end{figure}

The architecture for the system that takes care of quote generation and
job execution according to promised contracts can be seen in Figure~\ref{fig:architecture}.

Apart from separating simulation and real use clearly, we also split up our components in a back-end and front-end to allow
the pricing function to be implemented on a different platform than OpenStack. Our current implementation takes advantage of this
separation by implementing a revenue-optimizing function (see Section~\ref{sec:experiments}) in MatLab running on a Windows server. Both the front-end and back-end
interfaces are available as native python APIs and well as remote JSON/REST APIs.

The interaction described in Section~\ref{sec:pricemodel} 
is implemented in the following way:\\
The {\it Simulator} (or real user) creates a job according to the trace
specification (or user preferences) and asks the {\it QoS Manager} for a completion time price
quote (1). It in turn, via the {\it Node Allocator} (2) consults the
{\it Completion Time Predictor} to obtain a prediction (3) and
the {\it WTP Learner} for the assumed willingness-to-pay and deadline
distributions (see Section~\ref{sec:experiments}) (4), before asking the {\it Revenue Scheduler} for the actual quote (5).
Now the Simulator (or user) evaluates the quote and picks at most one contract (6)
according to an internal (not revealed to the system) willingness to pay
vs price surplus. Once a contract is selected the job can be scheduled through the
system via the {\it Job Executor} (7) into the Sahara OpenStack execution environment.

\section{Pricing Implementation and Experiments}\label{sec:experiments}
The goal of the experiments is to compare revenue of our proposed result-based selling mechanism to that of a resource-based pricing scheme
under different degrees of resource contention for a series of realistic job submissions.

Consider a cloud provider serving multiple customer types $k=1 \ldots K$. Customer types differ in their sensitivity to job completion time.
Specifically, assume a customer of
type $k$ has willingness to pay $W^k(t)$ for completion time $t \in [1,D^k]$, given by the function $W^k(t) = (D^k-t)^+ M^k/(D^k-1)$
where $M^k$ is uniformly distributed in $[m^k_l, m^k_u]$ and $D^k$ is uniformly distributed in $[d^k_l, d^k_u]$. This willingness to pay function
implies that the type-$k$ customer has maximum willingness to pay $M^k$ for completion time $1$, is willing to pay nothing for jobs finished beyond deadline $D^k$, and has linearly decreasing willingness to pay in the interval $[1,D^k]$.

When a job arrives, the provider does not know the type $k$ of the arriving customer, but he knows the distribution over customer types.
Let $\pi_k$ be the probability that a customer is of type $k$, where  $\sum_k{\pi_k} = 1$. For a given customer type,
the provider does not know the exact willingness to pay function of the customer but knows its functional form
and the distributions of $M^k$ and $D^k$. The provider also knows the two predicted possible completion times $t_{f}$ and $t_{s}$  ($t_{f} < t_{s}$) for the job,
based on the current system load.

In the results-based mechanism, the provider must choose prices $\textbf{\emph{p}}= (p_{f},p_{s})$  to offer the customer for completion times  $\textbf{\emph{t}}= (t_{f},t_{s})$,
respectively. We propose a pricing approach in which the provider sets prices to maximize his expected revenue from this job.
Let $q_f^k(\textbf{\emph{t}},\textbf{\emph{p}}) = P(W^k(t_{f})\ge \mbox{max} \{p_{f} ,  W^k (t_{s} )-p_{s} + p_{f} \} )$ represent the probability that the customer of type $k$ is willing to
pay at least $p_{f}$ for completion time $t_{f}$ and his surplus $W^k (t_{f} )-p_{f}$ for the fast completion time is at least as great as his surplus
$W^k (t_{s} )-p_{s}$ for the slow completion time. Similarly, let $q_s^k(\textbf{\emph{t}},\textbf{\emph{p}}) = P(W^k (t_{s})\ge \mbox{max} \{p_{s} , W^k (t_{f} )-p_{f}+p_{s} \})$ be the
probability that a type-$k$ customer prefers to purchase slow service at price $p_{s}$ to purchasing fast service at price $p_{f}$ or to purchasing nothing. Then the provider's expected revenue from this job, given prices  $\textbf{\emph{p}}= (p_{f},p_{s})$  for completion times  $\textbf{\emph{t}}= (t_{f},t_{s})$,  is $g(\textbf{\emph{t}},\textbf{\emph{p}})$,  given by:
\begin{equation}
g(\textbf{\emph{t}},\textbf{\emph{p}}) = \sum_{k=1}^{K} \pi_k \{ p_{f} q_f^k(\textbf{\emph{t}},\textbf{\emph{p}}) + p_{s} q_s^k(\textbf{\emph{t}},\textbf{\emph{p}}) \}
\end{equation}

This pricing method is myopic~\footnote{in that it chooses prices to maximize revenue from the current job without consideration of future jobs}. 
For demand-anticipating algorithms, we refer the reader to~\cite{bwh2015}.

 The resource-based pricing scheme we use as a benchmark, {\it Bench}, applies a single fixed price per node-period to all jobs. This fixed price is determined based on the same assumptions about the shape of the customers' willingness to pay function, and the distribution of the function's parameters. However, the fixed resource-based pricing model differs from the QoS based model in terms of the provider's knowledge of each specific job requirements and completion times when setting prices. Indeed, in the fixed price model, the provider must choose the fixed price independently of any particular job.
In the fixed price model, the interaction between the provider and the (potential) customer can be described according to the following protocol. First, the provider sets the per-node-period price. Given the price, the customer with job type $h$ decides what to buy. There are three possibilities: the customer buys $c_{f}^h$ nodes for $t_{f}^h$ periods, the customer buys $c_{s}^h$ nodes for $t_{s}^h$ periods, or the customer does not buy any nodes.
Both the job resource requirements and his willingness to pay are the customer's private information. Instead, the provider knows the distribution $\gamma = (\gamma_{h})$ over job types $h$, where for each possible job type $h$ she knows the node requirements in each period $c_{f}^h$ and $c_{s}^h$ and job durations $t_{f}^h$ and $t_{s}^h$ for fast and slow options. We enforce that a job is only accepted if it can be scheduled without delay, so that the job durations $t_{f}^h$ and $t_{s}^h$ correspond to completion times.

We note that there is an admission control feature built into the system
that rejects the job submission if no price quotes could be given (because resources are not available to start the job immediately) or if the
given quotes are too expensive for the user, based on their preferences.

The experimental design is to vary supply and demand and compare the performance of our QoS-based pricing
algorithm against {\it Bench}, the baseline resource-based pricing scheme.
We vary supply by considering different numbers of nodes in the cluster, and vary demand by considering a range of different mean interarrival times between jobs.

A job trace simulator generates traces based on uniform distributions of max willingness-to-pay
and deadline parameters within the ranges shown in Table~\ref{T:wtpconfig}.
Deadlines are specified in minutes.

\begin{table}[h!]
\centering
\begin{tabular}{|c|c|c|}
 \hline
 Customer Type & Max WTP Range & Deadline Range \\
 \hline\hline
 1 & [100,120] & [20,30] \\
 2 & [45,55] & [60.70] \\
 \hline
\end{tabular}
\vspace{0.5cm}
\caption{Willingness-to-pay preferences across the two customer types used in the experiments.}
\label{T:wtpconfig}
\end{table}

Each job is mapped randomly to one of the 50 tenants~\footnote{A tenant here is an OpenStack project namespace with separate virtual networks, storage and clusters}
with equal probability, and each tenant is
assigned to one of the customer types with equal probability.

The trace also comprises two types of jobs: an IO-intensive job and a CPU-intensive job.
The IO-intensive job is the standard Hadoop WordCount example using an input corpus of 10GB
text~\footnote{The COCA sample from http://corpus.byu.edu/full-text/ duplicated with a factor of 10000}
stored in compressed form (3.5GB) in each tenant's own OpenStack Swift container separately. The
data is pulled from Swift and decompressed into HDFS when the job is submitted. At the end of the
computation the resulting output (single reducer) is copied back to Swift. The CPU-intensive
job is the Hadoop Pi example, using a Monte Carlo simulation to find decimals of Pi at different
levels of accuracy controlled by task and sample counts. We used 200 tasks and 5000 samples
for the jobs in our experiment. The completion times for these jobs were predicted out-of-band for
different node configurations. See Table~\ref{T:jobpredict}. The clusters all use the single-VCPU,
small instance flavor of OpenStack.

\begin{table}[h!]
\centering
\begin{tabular}{|c|c|c|}
 \hline
 Job type & 3-node completion & 10-node completion \\
 \hline\hline
 IO & 51 & 23 \\
 CPU & 9 & 5 \\
 \hline
\end{tabular}
\vspace{0.5cm}
\caption{Predicted completion times in minutes.}
\label{T:jobpredict}
\end{table}

Now, we use this configuration to generate traces of job submissions as a Poisson
process with interarrival time means of 20, 25, 30, 35 and 40 seconds. Each trace
with a specific interarrival time (IAT) configuration is then run through the system with different numbers of nodes in the cluster. The capacity levels used were 50, 75, 100, 125 and 150 nodes.
A job submission is not be admitted if there are not enough available nodes to place the job, i.e.
no queuing is done. In this case the pricing functions will not generate a price menu.

The experiments track metrics such as revenue, net utility (user willingness to pay minus price paid),
utilization, admission rate, successful contracts, as well as other low level system parameters for verification.

\begin{figure}[htp]
\vspace{-0.3cm}
\subfigure[Revenue gain.]{
\hspace{-0.5cm}
\includegraphics[scale=0.17]{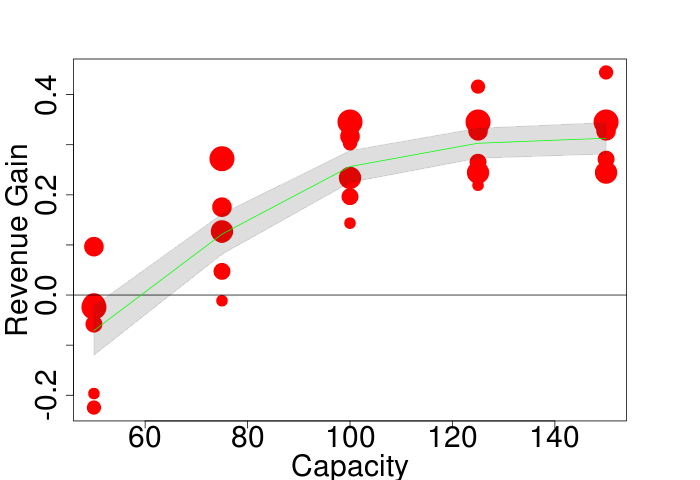}
\label{fig:metricCapacity}}
\subfigure[Utility gain.]{
\hspace{-0.5cm}
\includegraphics[scale=0.17]{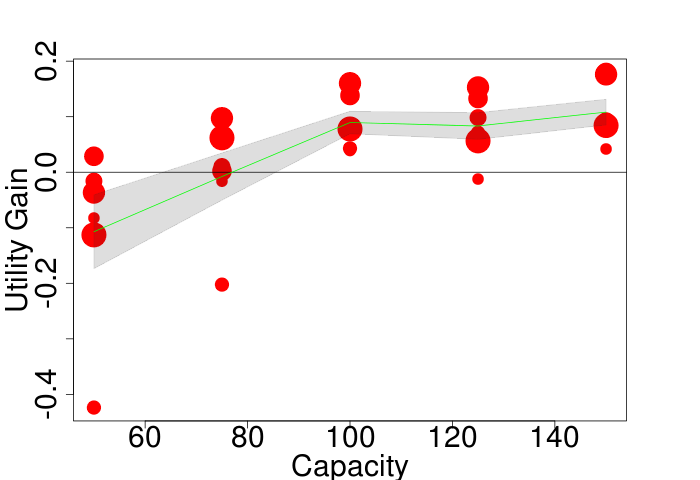}
\label{fig:metricUtility}}
\subfigure[Revenue under Contention.] {
\hspace{-0.5cm}
\includegraphics[scale=0.17]{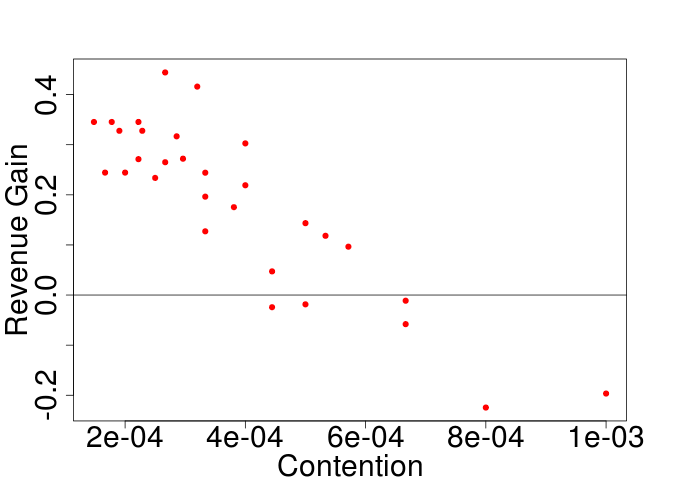}
\label{fig:contention}}
\subfigure[Admission Rate gain.]{
\hspace{-0.5cm}
\includegraphics[scale=0.17]{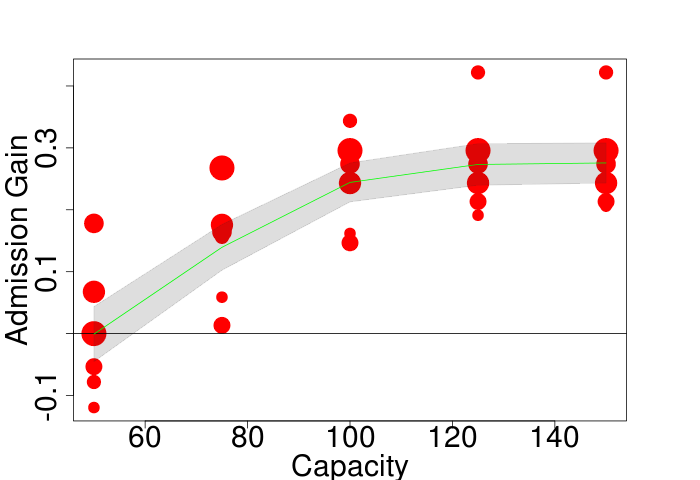}
\label{fig:metricAdmission}}
\subfigure[Utilization gain.]{
\hspace{-0.5cm}
\includegraphics[scale=0.17]{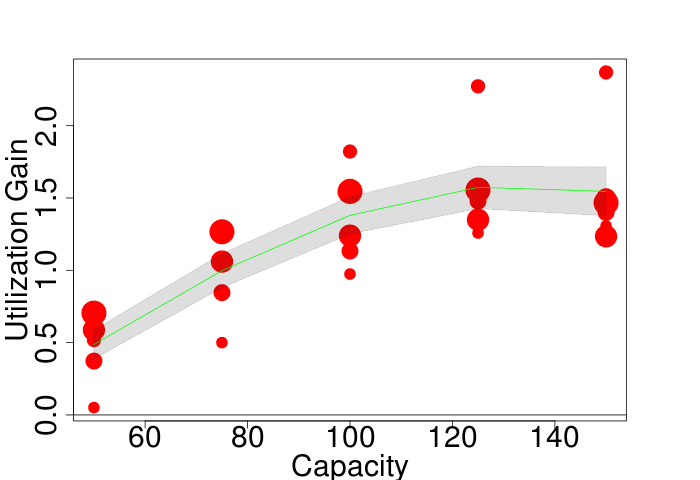}
\label{fig:metricUtilization}}
\subfigure[Gain in successful jobs.]{
\includegraphics[scale=0.17]{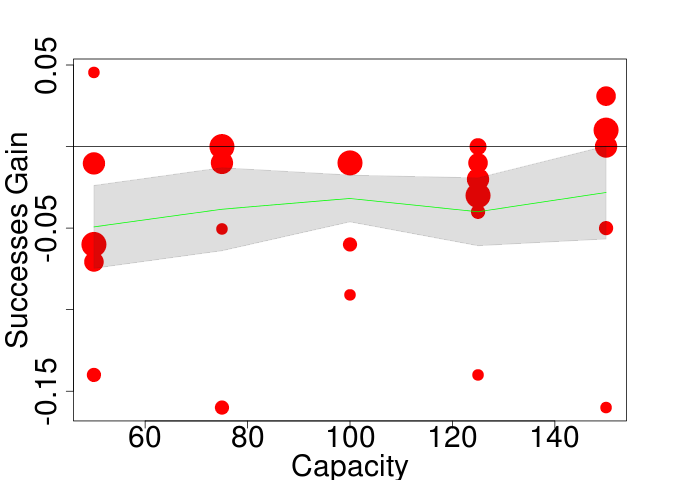}
\label{fig:metricSuccesses}}
\label{fig:experimentresult}
\caption{Experiment result under different node capacity and IAT settings.
Larger red dots represent longer IAT (with 20,25,30,35,40, and 45 seconds). 
The gray area 
represents one standard error variation across all IAT settings.}
\vspace{-0.5cm}
\end{figure}

We show the revenue gain across capacity and IAT configurations
in Figure~\ref{fig:metricCapacity}, the net utility gain, Figure~\ref{fig:metricUtility}, as well as the
revenue at different levels of contention $(IAT \cdot Capacity)^{-1}$, Figure~\ref{fig:contention}.
We observe that there is a revenue gain from our algorithm across all configurations and that it is
up to 40\% higher than the resource-based {\it Bench} pricing scheme.
It is interesting to note that
although we do not optimize for customer net utility (satisfaction), and our QoS-based pricing algorithm does not know the
utility function of the user submitting a job (just the overall distribution of utility functions), we improve the
aggregate utility in the system by up to 20\%. 
When studying the revenue gain based on the contention in the system, it is clear that higher contention leads to less opportunity to gain revenue. This happens because of the myopia of the current algorithm; 
it does not take future expected demand into account when admitting and pricing jobs. 
This finding suggests simple heuristic for improving revenue: add a price premium that increases with the degree of contention. 
Then the provider would admit fewer jobs and reserve capacity for future high-revenue jobs. 

To understand better why we achieve this improvement we also look at the admission rate and utilization
in Figure~\ref{fig:metricAdmission} and Figure~\ref{fig:metricUtilization}, respectively.
It is clear from
these graphs that the higher revenue and utility extracted from our algorithm is due to admitting more jobs and making better
use of available resources.

The downside of admitting more jobs and creating more load and competition in the system is that it also increases the risk of interference
between jobs, causing missed deadlines~\footnote{which we don't adress in this work, i.e we assume that we are conservative enough in the prediction to avoid them}. 
To study this effect we look at the contract success rate, defined as the fraction of scheduled jobs that are completed within their promised time.
Figure~\ref{fig:metricSuccesses} shows that the contract success rate under QoS-based pricing is within 5-15\% of that of the baseline. Missed deadlines increase with the degree of contention as defined above. Pricing algorithms may factor this effect in by explicitly including a likelihood for contract failure
based on load in the revenue optimization model. Alternatively, one could add a premium to the price under
heavy load to compensate for the increased likelihood of breaking contracts. 
Our future work is focused on addressing the effects of contention by taking expected future demand into account in the pricing algorithm.

\begin{table}[h!]
\centering
\begin{tabular}{|l|c|c|}
 \hline
 & Bench & RevOp \\
 \hline\hline
 Revenue & $2929\pm161$ & $3471\pm523$ \\
 Utility & $1596\pm154$ & $1648\pm245$ \\
 Utilization & $0.32\pm.15$ & $\mathbf{0.65\pm0.14}$ \\
 Node Periods & $834\pm75$ & $\mathbf{3504\pm867}$ \\
 Success Rate & $0.99\pm0.02$ & $0.95\pm0.05$ \\
 Max Load & $50\pm9$ & $\mathbf{95\pm31}$ \\
 Admission Rate & $0.71\pm0.04$ & $0.84\pm0.10$ \\
 \hline
\end{tabular}
\vspace{0.5cm}
\caption{Summary of Results with 95\% confidence bounds.}
\label{T:summaryresults}
\end{table}

\section{Conclusions}\label{sec:conclusions}
We have shown that a simple revenue-optimizing pricing function can improve the 
revenue by 40\% compared to a per-node fixed pricing model for batch jobs running
in a multi-tenant OpenStack Sahara environment. We see the best results under low
contention, as the price-discrimination-based scheduler can admit more jobs.  For the
same reason our method also improves social welfare, or utility across users under
low contention. This result is promising as it shows that our method can make good use 
of idle resources in the Cloud by improving utilization during low demand.

Future work includes providing an on-line predictor of completion times, and applying
more sophisticated pricing algorithms, e.g. that take future demand into account~\cite{bwh2015}. We are also working on 
a tighter integration with OpenStack and Sahara to support a wider range of batch applications.

\bibliographystyle{abbrvnat}
\bibliography{related}
\end{document}